\input harvmac
\def\gsim{{~\raise.15em\hbox{$>$}\kern-.85em
          \lower.35em\hbox{$\sim$}~}}
\def\lsim{{~\raise.15em\hbox{$<$}\kern-.85em
          \lower.35em\hbox{$\sim$}~}}
\def\Re{{\cal R}e}
\def\Im{{\cal I}m}
\def\barD{\overline{D^0}}
\def\DDbar{D^0-\barD}
\def\BR{{\rm BR}}

\noblackbox
\baselineskip 18pt plus 2pt minus 2pt
\Title{\vbox{\baselineskip12pt
\hbox{hep-ph/0005181}
\hbox{WIS/8/00-May-DPP}
\hbox{SLAC-PUB-8457}
\hbox{FERMILAB-Pub-00/102-T}
\hbox{IASSNS--HEP--00--42}
\hbox{JHU-TIPAC-200003}
\vskip -1.5truecm
}}
{\vbox{ 
\centerline{Lessons from CLEO and FOCUS Measurements}
\smallskip
\centerline{of $\DDbar$ Mixing Parameters}
} }
\vskip -1truecm
\centerline{S. Bergmann$^a$, Y. Grossman$^b$, Z. Ligeti$^c$, 
Y. Nir$^{d,a}$ and A.A. Petrov$^e$}
\medskip
\centerline{\it $^a$Department of Particle Physics,
Weizmann Institute of Science, Rehovot 76100, Israel}
\smallskip
\centerline{\it $^b$Stanford Linear Accelerator Center, 
Stanford University, Stanford, CA 94309, USA\foot{
Research supported
by the Department of Energy under contract DE-AC03-76SF00515.}}
\smallskip
\centerline{\it $^c$Theory Group, Fermilab, P.O. Box 500,
 Batavia, IL 60510, USA}
\smallskip
\centerline{\it $^d$School of Natural Sciences,
Institute for Advanced Study, Princeton, NJ 08540, USA\foot{
Address for academic year 1999-2000.}}
\smallskip
\centerline{\it $^e$\ Department of Physics and Astronomy,
 The Johns Hopkins University}
\centerline{\it 3400 North Charles Street, Baltimore, MD 21218, USA}
\bigskip

\baselineskip 18pt
\noindent
If the true values of the $\DDbar$ mixing parameters lie within
the one sigma ranges of recent measurements, then there is strong
evidence for a large width difference, $y\gsim0.01$, and large
$SU(3)$ breaking effects in strong phases, $\delta\gsim\pi/4$.
These constraints are model independent, and would become stronger
if $|M_{12}/\Gamma_{12}|\ll1$ in the $\DDbar$ system.
The interesting fact that the FOCUS result cannot be explained by a large
mass difference 
is not trivial and depends on the small $D^0/\barD$ production asymmetry
in FOCUS and the bounds on CP violating effects from CLEO.
The large value of $\delta$ might help explain why $y\sim\sin^2\theta_c$.

\vfill

\Date{5/00}

\newsec{Introduction}
\nref\cleoexp{R.~Godang {\it et al.} [CLEO Collaboration], hep-ex/0001060.}%
\nref\focusexp{J.M. Link {\it et al.} [FOCUS Collaboration], hep-ex/0004034.}%
Recent studies of time-dependent decay rates of $D^0\rightarrow K^+\pi^-$ 
by the CLEO collaboration \cleoexp\ and measurements of the combination of 
$D^0\rightarrow K^+K^-$ and $D^0\rightarrow K^-\pi^+$ rates by the FOCUS
collaboration \focusexp\ have provided highly interesting results
concerning $\DDbar$ mixing. (For previous, related results, see
\nref\esixno{J.C. Anjos {\it et al.} [E691 Collaboration],
 Phys. Rev. Lett. 60 (1988) 1239.}%
\nref\cleoa{D.~Cinabro {\it et al.} [CLEO Collaboration],
 Phys. Rev. Lett. 72 (1994) 1406.}%
\nref\esnosl{E.M.~Aitala {\it et al.} [E791 Collaboration],
 Phys. Rev. Lett. 77 (1996) 2384, hep-ex/9606016.}%
\nref\esnoa{E.M.~Aitala {\it et al.} [E791 Collaboration],
 Phys. Rev. D57 (1998) 13, hep-ex/9608018.}%
\nref\aleph{R.~Barate {\it et al.} [ALEPH Collaboration],
 Phys. Lett. B436 (1998) 211, hep-ex/9811021.}%
\nref\esnob{E.M.~Aitala {\it et al.} [E791 Collaboration],
 Phys. Rev. Lett. 83 (1999) 32, hep-ex/9903012.}%
\refs{\esixno-\esnob}.) Each of the two experiments
finds a signal for mixing at a level that is close to
$2\sigma$. It is not unlikely that these signals are
just the results of statistical fluctuations and the
true mixing parameters lie well below the experimental
sensitivity. In this work, however, we interpret the
experimental results assuming that their central values
are not far from the true values and that $\DDbar$ mixing
has indeed been observed. 

\nref\FNP{A.F. Falk, Y. Nir and A.A. Petrov,
 JHEP 12 (1999) 019, hep-ph/9911369.}%
The interpretation of the results and, in particular, testing
the consistency of the two recent measurements with each other,
require a careful treatment of signs and phase conventions.
We present the relevant model-independent formalism in section 2.
In section~3 we carefully explain what parameters have the FOCUS and CLEO
experiments actually measured. We emphasize that, in principle, 
both CLEO and FOCUS results can be accounted for even if the width 
difference is negligibly small. This fact was known for the CLEO 
result~\FNP, but it is much more subtle for the FOCUS result. 

In section 4, we analyze the theoretical implications of the FOCUS and 
CLEO results in a model independent framework. We do however make some 
reasonable assumptions. With new physics, it is possible that there are 
large, CP violating new contributions to the mass difference. On the 
other hand, it is very unlikely that the width difference 
\ref\BSN{G. Blaylock, A. Seiden and Y. Nir, 
 Phys. Lett. B355 (1995) 555, hep-ph/9504306.}\
and relevant decay amplitudes
\ref\BeNi{S. Bergmann and Y. Nir, JHEP 09 (1999) 031, hep-ph/9909391.}\ 
are significantly affected by new physics. In such a framework, the measured 
observables depend on the mass difference $x$, the width difference $y$, 
two independent CP violating parameters, $\phi$ and $A_m$, and a strong 
phase $\delta$. We find that the experimental results have strong implications
for the width difference $y$ and for the strong phase $\delta$.
The qualitative features are independent of the other parameters,
though the detailed quantitative results are not. 

It could be that the $\DDbar$ system is a unique example of
a case where the dispersive part of the $D^0\rightarrow\barD$ 
transition amplitude is much smaller than the absorptive
part, $|M_{12}|\ll|\Gamma_{12}|$. (For the $K^0-\overline{K^0}$
the two are comparable, while for the $B^0-\overline{B^0}$ and
$B_s-\overline{B_s}$ systems the situation is opposite,
$|M_{12}|\gg|\Gamma_{12}|$.) This situation, which is rarely discussed
in the literature, is analyzed in section 5. We point out that, if this 
approximation is valid, the dependence on $x$ and on the CP violating 
parameters can be neglected. Consequently, the FOCUS and CLEO results depend on 
$y$ and $\delta$ only, and the implications become much clearer, both 
qualitatively and quantitatively.

Within the Standard Model, $\DDbar$ mixing vanishes in the limit
of exact $SU(3)$ flavor symmetry of the strong interactions.
For example, the sum of the contributions to the width difference
from intermediate $K^+K^-$, $\pi^+\pi^-$, $K^+\pi^-$ and
$K^-\pi^+$ states vanishes in the $SU(3)$ limit. The fact that
the one sigma ranges of the FOCUS and CLEO results constrain
$\cos\delta$ allows, for the first time, a calculation of this
contribution based entirely on experimental data. We carry out such
a calculation in section 6 and find a surprisingly large contribution 
to $y$, of order one percent.

A summary of our results is given in section 7.

\newsec{Notations and Formalism}
We investigate neutral $D$ decays. The two mass eigenstates,
$|D_1\rangle$ of mass $m_1$ and width $\Gamma_1$ and
$|D_2\rangle$ of mass $m_2$ and width $\Gamma_2$,
are linear combinations of the interaction eigenstates:
\eqn\masint{\eqalign{
|D_1\rangle\ =&\ p|D^0\rangle+q|\barD\rangle,\cr
|D_2\rangle\ =&\ p|D^0\rangle-q|\barD\rangle.\cr}}
The average mass and width are given by
\eqn\SumMG{m\equiv {m_1+m_2\over2},\ \ \ 
\Gamma\equiv{\Gamma_1+\Gamma_2\over2}.}
The mass and width difference are parametrized by
\eqn\DelMG{x\equiv{m_2-m_1\over\Gamma},\ \ \ 
y\equiv{\Gamma_2-\Gamma_1\over2\Gamma}.}
Decay amplitudes into a final state $f$ are defined by
\eqn\AbarA{A_f\equiv\vev{f|{\cal H}_d|D^0},\ \ \ 
\bar A_f\equiv\vev{f|{\cal H}_d|\barD}.}
It is useful to define the complex parameter $\lambda_f$:
\eqn\deflam{\lambda_f\equiv{q\over p}\ {\bar A_f\over A_f}.}

The processes that are relevant to the CLEO and FOCUS experiments
are the doubly-Cabibbo-suppressed $D^0\rightarrow K^+\pi^-$ decay,
the singly-Cabibbo-suppressed $D^0\rightarrow K^+K^-$ decay,
the Cabibbo-favored $D^0\rightarrow K^-\pi^+$ decay,
and the three CP-conjugate decay processes. We now write down approximate 
expressions for the time-dependent decay rates that are valid for times 
$t\lsim1/\Gamma$. We take into account the experimental information that $x$, 
$y$ and $\tan\theta_c$ are small, and expand each of the rates only to the
order that is relevant to the CLEO and FOCUS measurements:
\eqn\dpikt{\eqalign{
\Gamma[D^0(t)\rightarrow&\ K^+\pi^-]\ =\ e^{-\Gamma t}|\bar A_{K^+\pi^-}|^2
|q/p|^2\cr \times&\left\{|\lambda_{K^+\pi^-}^{-1}|^2+
[\Re(\lambda^{-1}_{K^+\pi^-})y+\Im(\lambda^{-1}_{K^+\pi^-})x]\Gamma t
+{1\over4}(y^2+x^2)(\Gamma t)^2\right\},\cr
\Gamma[\barD(t)\rightarrow&\ K^-\pi^+]\ =\ e^{-\Gamma t}|A_{K^-\pi^+}|^2
|p/q|^2\cr \times&\left\{|\lambda_{K^-\pi^+}|^2+[\Re(\lambda_{K^-\pi^+})y
+\Im(\lambda_{K^-\pi^+})x]\Gamma t+{1\over4}(y^2+x^2)(\Gamma t)^2\right\},\cr}}
\eqn\dkkt{\eqalign{
\Gamma[D^0(t)\rightarrow K^+K^-]\ =&\ e^{-\Gamma t}|A_{K^+K^-}|^2
\left\{1+[\Re(\lambda_{K^+K^-})y-\Im(\lambda_{K^+K^-})x]\Gamma t\right\},\cr
\Gamma[\barD(t)\rightarrow K^+K^-]\ =&\ e^{-\Gamma t}|\bar A_{K^+K^-}|^2
\left\{1+[\Re(\lambda^{-1}_{K^+K^-})y-\Im(\lambda^{-1}_{K^+K^-})x]
\Gamma t\right\},\cr}}
\eqn\dkpit{\eqalign{
\Gamma[D^0(t)\rightarrow&\ K^-\pi^+]\ =\ e^{-\Gamma t}|A_{K^-\pi^+}|^2,\cr
\Gamma[\barD(t)\rightarrow&\ K^+\pi^-]\ =\ 
e^{-\Gamma t}|\bar A_{K^+\pi^-}|^2.\cr}}

Within the Standard Model, the physics of $\DDbar$ mixing and of the
tree level decays is dominated by the first two generations and,
consequently, CP violation can be safely neglected. In all `reasonable'
extensions of the Standard Model, the six decay modes of eqs. \dpikt,
\dkkt\ and \dkpit\ are still dominated by the Standard Model CP conserving 
contributions \BeNi. On the other hand, there could be new short distance, 
possibly CP violating contributions to the mixing amplitude $M_{12}$.
Allowing for only such effects of new physics, the picture of CP violation
is simplified since there is no direct CP violation. The effects of indirect
CP violation can be parametrized in the following way
\ref\nirssi{Y. Nir, hep-ph/9911321.}:
\eqn\parlkpi{\eqalign{
|q/p|\ =&\ R_m,\cr
\lambda^{-1}_{K^+\pi^-}\ =&\ \sqrt{R}\ R_m^{-1}\ e^{-i(\delta+\phi)},\cr
\lambda_{K^-\pi^+}\ =&\ \sqrt{R}\ R_m\ e^{-i(\delta-\phi)},\cr
\lambda_{K^+K^-}\ =&\ -R_m\ e^{i\phi}.\cr}}
Here $R$ and $R_m$ are real and positive dimensionless numbers.
CP violation in mixing is related to $R_m\neq1$ while CP violation in the 
interference of decays with and without mixing is related to $\sin\phi\neq0$. 
The choice of phases and signs in \parlkpi\ is consistent with having $\phi=0$ 
in the Standard Model and $\delta=0$ in the $SU(3)$ limit (see below). 
We further define
\eqn\defxy{\eqalign{
x^\prime\ \equiv&\ x\cos\delta+y\sin\delta,\cr
y^\prime\ \equiv&\ y\cos\delta-x\sin\delta.\cr}}

With our assumption that there is no direct CP violation in the processes
that we study, and using the parametrizations \parlkpi\ 
and \defxy, we can rewrite eqs. \dpikt$-$\dkpit\ as follows:
\eqn\dpikcpv{\eqalign{
\Gamma[D^0(t)\rightarrow&\ K^+\pi^-]\ =\ e^{-\Gamma t}|A_{K^-\pi^+}|^2\cr 
 \times&\left[R+\sqrt{R}R_m(y^\prime\cos\phi-x^\prime\sin\phi)\Gamma t
 +{R_m^2\over4}(y^2+x^2)(\Gamma t)^2\right],\cr
\Gamma[\barD(t)\rightarrow&\ K^-\pi^+]\ =\ e^{-\Gamma t}|A_{K^-\pi^+}|^2\cr
 \times&\left[R+\sqrt{R}R_m^{-1}(y^\prime\cos\phi+x^\prime\sin\phi)
 \Gamma t+ {R_m^{-2}\over4}(y^2+x^2)(\Gamma t)^2\right]\cr}}
\eqn\dkkcpv{\eqalign{ 
\Gamma[D^0(t)\rightarrow&\ K^+K^-]\ =\ e^{-\Gamma t}|A_{K^+K^-}|^2
\left[1-R_m(y\cos\phi-x\sin\phi)\Gamma t\right],\cr
\Gamma[\barD(t)\rightarrow&\ K^+K^-]\ =\ e^{-\Gamma t}|A_{K^+K^-}|^2
\left[1-R_m^{-1}(y\cos\phi+x\sin\phi)\Gamma t\right],\cr}}
\eqn\dkpicpv{
\Gamma[D^0(t)\rightarrow K^-\pi^+]\ =\ 
\Gamma[\barD(t)\rightarrow K^+\pi^-]\ =\ e^{-\Gamma t}|A_{K^-\pi^+}|^2.}

\newsec{CLEO and FOCUS Measurements}
The FOCUS experiment \focusexp\ fits the time dependent decay rates of the 
singly-Cabibbo suppressed \dkkcpv\ and the Cabibbo-favored \dkpicpv\ modes to 
pure exponentials. We define $\hat\Gamma$ to be the parameter that is 
extracted in this way. More explicitly, for a time dependent decay rate with
$\Gamma[D(t)\rightarrow f]\propto e^{-\Gamma t}(1-z\Gamma t+\cdots)$, where
$|z|\ll1$, we have $\hat\Gamma(D\rightarrow f)=\Gamma(1+z)$.
The above equations imply the following relations:
\eqn\fitexp{\eqalign{
\hat\Gamma(D^0\rightarrow K^+K^-)\ =&\
 \Gamma\ [1+R_m(y\cos\phi-x\sin\phi)],\cr
\hat\Gamma(\barD\rightarrow K^+K^-)\ =&\
 \Gamma\ [1+R_m^{-1}(y\cos\phi+x\sin\phi)],\cr
\hat\Gamma(D^0\rightarrow K^-\pi^+)\ =&\
 \hat\Gamma(\barD\rightarrow K^+\pi^-)\ =\ \Gamma.\cr}}
Note that deviations of $\hat\Gamma(D\rightarrow K^+K^-)$ from $\Gamma$ 
do not require that $y\neq0$. They can be accounted for by $x\neq0$ and 
$\sin\phi\neq0$, but then they have a different sign in the $D^0$ and $\barD$ 
decays. FOCUS combines the two $D\rightarrow K^+K^-$ modes. To understand the 
consequences of such an analysis, one has to consider the relative
weight of $D^0$ and $\barD$ in the sample. Let us define 
$A_{\rm prod}$ as the production asymmetry of $D^0$ and $\barD$:
\eqn\defaprod{A_{\rm prod}\equiv{N(D^0)-N(\barD)\over N(D^0)+N(\barD)}.}
Then
\eqn\FkkGcpv{\eqalign{y_{\rm CP}\ \equiv&\ {\hat\Gamma(D\rightarrow K^+K^-)\over
\hat\Gamma(D^0\rightarrow K^-\pi^+)}-1\cr\
 =&\ y\cos\phi\left[{1\over2}(R_m+R_m^{-1})+{A_{\rm prod}\over2}(R_m-R_m^{-1})\right]\cr
 &-x\sin\phi\left[{1\over2}(R_m-R_m^{-1})+{A_{\rm prod}\over2}(R_m+R_m^{-1})\right].\cr}}
The one sigma range measured by FOCUS is
\eqn\ycpnum{y_{\rm CP}=(3.42\pm1.57)\times10^{-2}.}

The interpretation of this measurement simplifies when the following two facts
are taken into account:
\item{(i)} The E687 data 
\ref\eses{P.L. Frabetti {\it et al.} [E687 Collaboration],
 Phys. Lett. B370 (1996) 222.}\
suggest that $A_{\rm prod}$ is small for FOCUS, of order 0.03.
\item{(ii)} The CLEO data \cleoexp\ suggest that $R_m$ is not very different 
from one (see below). Actually, CLEO implicitly assume that this is the case 
in their analysis by using
\eqn\defAm{R_m^{\pm2}=1\pm A_m.}

Evaluating \FkkGcpv\ to linear order in the small quantities $A_{\rm prod}$
and $A_m$ yields
\eqn\FkkGcp{y_{\rm CP}=y\cos\phi-x\sin\phi\left({A_m\over2}+A_{\rm prod}\right).}

The CLEO measurement \cleoexp\ gives the coefficient of each of the three 
terms ($1$, $\Gamma t$ and $(\Gamma t)^2$) in the doubly-Cabibbo suppressed
decays \dpikcpv. Such measurements allow a fit to the parameters $R$, $R_m$, 
$x^\prime\sin\phi$, $y^\prime\cos\phi$, and $x^2+y^2$.  Fit A of ref. 
\cleoexp\ quotes the following one sigma ranges:\foot{CLEO quote a range for 
$y^\prime$. It is obvious however that, with our conventions, their range 
applies to $y^\prime{\rm sign}(\cos\phi)$ or perhaps to $y^\prime\cos\phi$. 
Since the one sigma range is $|\cos\phi|\gsim0.8$, the difference between these
two possibilities is unimportant for our purposes.}
\eqn\cleoyx{\eqalign{
R\ =&\ (0.48\pm0.13)\times10^{-2},\cr
y^\prime\cos\phi\ =&\ (-2.5^{+1.4}_{-1.6})\times10^{-2},\cr
x^\prime\ =&\ (0.0\pm1.5)\times10^{-2},\cr
A_m\ =&\ 0.23^{+0.63}_{-0.80}.\cr}}

\nref\bartelt{J. Bartelt {\it et al.} [CLEO Collaboration],
 Phys. Rev. D52 (1995) 4860.}%
\nref\esnoc{E.M.~Aitala {\it et al.} [E791 Collaboration],
 Phys. Lett. B421 (1998) 405, hep-ex/9711003.}%
\nref\Wiss{J. Wiss, Fermilab Joint Experimental Theoretical Seminar 
(April 2000), available from 
http://www-hep.colorado.edu/FocusPublic/ConferencesPublic.}%
\nref\cleokk{D. Asner, T. Hill, S. McGee and C. Prescott
[CLEO Collaboration],
Presented at APS April Meeting 2000, Long Beach, CA.}%
We would like to point out that the interpretation of the FOCUS
and CLEO results in terms of $y$, $x$, $\phi$, $\delta$ and $A_m$ is
almost independent of our assumption that there is no CP violation
in decay. To understand this point, let us parametrize CP violation
in decay in the following way:
\eqn\defadec{\eqalign{A_{\rm CP}(f)\ \equiv&\ {
\Gamma(D^0\rightarrow f)-\Gamma(\barD\rightarrow\bar f)\over
\Gamma(D^0\rightarrow f)+\Gamma(\barD\rightarrow\bar f)}\cr
=&\ {1-|\bar A_{\bar f}/A_f|^2\over1+|\bar A_{\bar f}/A_f|^2}.\cr}}
Experimentally, we have the following constraints on the asymmetries
in the Cabibbo-favored \bartelt, singly-Cabibbo-suppressed 
\refs{\esnoc-\cleokk}\
and doubly-Cabibbo-suppressed \cleoexp\ decays:
\eqn\expacp{\eqalign{
A_{\rm CP}(K^-\pi^+)\ =&\ 0.001\pm0.011,\cr
A_{\rm CP}(K^-K^+)\ =&\ 0.0004\pm0.0234,\cr
A_{\rm CP}(K^+\pi^-)\ =&\ -0.01\pm0.17.\cr}}
For FOCUS, eq. \FkkGcp\ would be corrected by terms of order 
$A_{\rm CP}(K^-K^+)A_{\rm prod}$ and $A_{\rm CP}(K^-K^+)A_m$,
which are negligible. For CLEO, the results in
eq. \cleoyx\ have been obtained allowing for CP violation in decay.
There is however another subtle aspect of direct CP violation
where our theoretical assumption does play a role. In the presence
of new CP violating contributions to the decay amplitudes, the CP
violating phases in $\lambda_f$ are not necessarily universal.
Therefore, the use of a single phase $\phi$ in eq. \parlkpi\ and
consequently in eqs. \FkkGcp\ and \cleoyx\ is valid only in the 
absence of direct CP violation.

\newsec{Theoretical Interpretation}
We now assume that the true values of the various mixing parameters
are within the one sigma ranges measured by FOCUS and CLEO. That means
in particular that we hypothesize that $\DDbar$ mixing is being
observed in the FOCUS measurement of $y_{\rm CP}$ and in the CLEO measurement
of $y^\prime\cos\phi$. The combination of these two results is particularly
powerful in its theoretical implications.

\nref\hnne{H.N. Nelson, hep-ex/9908021.}%
\nref\BiUr{I.I. Bigi and N.G. Uraltsev, hep-ph/0005089.}%
\nref\GLP{A.F. Falk, Y. Grossman, Z. Ligeti and A.A. Petrov, to appear.}%
Let us first focus on the FOCUS result \ycpnum. We argue that it is very unlikely
that this result is accounted for by the second term in \FkkGcp. Even if we take 
all the relevant parameters to be close to their one sigma upper bounds, say,
$|x|\sim0.04$  (we use Fig. 3 of ref. \cleoexp\ to extract this upper bound), 
$|\sin\phi|\sim0.6$, $|A_m/2|\sim0.4$ and $A_{\rm prod}\sim0.03$, we get
$y_{\rm CP}\sim0.01$, about a factor of two too small. We can make then the
following {\it model independent} statement: {\it if the true values of the 
mixing parameters are within the one sigma ranges of CLEO and FOCUS measurements, 
then $y$ is of order of a (few) percent.} Note that this is true even in the presence 
of CP violation, which does allow a mass difference, $x\neq0$, to mimic a deviation
from the average lifetime. Practically, we can take the FOCUS result to be given to
a good approximation by
\eqn\newfoc{y\cos\phi\approx0.034\pm0.016.}
This is a rather surprising result. Most theoretical estimates are well
below the one percent level (for a review, see \hnne).
These estimates have however been recently criticized \refs{\BiUr,\GLP}.
We will have more to say about this issue in section 6.

Second, we examine the consistency of the FOCUS and CLEO results. The two most
significant measurements, that of $y\cos\phi$ in eq. \newfoc\ and that of
$y^\prime\cos\phi$ in eq. \cleoyx\ are consistent if
\eqn\difocl{\cos\delta-(x/y)\sin\delta=-0.73\pm0.55.}
This requirement allows us to make a second {\it model independent} statement:
{\it if the true values of the mixing parameters are within the one sigma ranges 
of CLEO and FOCUS measurements, then the difference in strong phases between the 
$D^0\rightarrow K^+\pi^-$ and $D^0\rightarrow K^-\pi^+$ decays is very large.} 
For $\delta=0$ we get $y^\prime/y=1$ instead of the range given in eq. \difocl. 
To satisfy \difocl, we need, for example, 
\eqn\delxy{\cos\delta\lsim\cases{+0.65&$|x|\sim|y|$,\cr -0.18&$|x|\ll|y|$.\cr}}

The result in eq. \delxy\ is also rather surprising. The strong phase $\delta$
vanishes in the $SU(3)$ flavor symmetry limit
\ref\Wolf{L. Wolfenstein, Phys. Rev. Lett. 75 (1995) 2460, hep-ph/9505285.}.
None of the models in the literature
\nref\ChCh{L. Chau and H. Cheng, Phys. Lett. B333 (1994) 514, hep-ph/9404207.}%
\nref\BrPa{T.E. Browder and S. Pakvasa,
 Phys. Lett. B383 (1996) 475, hep-ph/9508362.}%
\refs{\FNP,\ChCh,\BrPa}\ finds such a large $\delta$. Eq. \delxy\ implies a very 
large $SU(3)$ breaking effect in the strong phase. For comparison, the experimental 
value of $\sqrt{R}\sim0.07$ in eq. \cleoyx\ is enhanced compared to its $SU(3)$ 
value of $\tan^2\theta_c\sim0.051$ by a factor $\sim1.4$. On the other hand,
there are other known examples of $SU(3)$ breaking effects of order one in
$D$ decays,\foot{For example, $\Gamma(D^0\rightarrow K^+K^-)
/\Gamma(D^0\rightarrow\pi^+\pi^-)=2.75\pm0.15\pm0.16$ experimentally 
\esnoc, while the ratio is predicted to be one in the $SU(3)$ limit.}\
so perhaps we should not be prejudiced against a very large $\delta$. 

Before concluding this section, we would like to explain the consequences 
of the CLEO and FOCUS measurements in the context of the Standard Model.
Within the Standard Model, $\DDbar$ mixing and $D^0$ decays into $K^+K^-$,
$\pi^+\pi^-$ and $\pi^\pm K^\mp$ are described to an excellent approximation
by physics of the first two generations. Consequently, the Standard Model
makes a clean prediction that any CP violating effects in these processes
are negligibly small. We can thus safely set $\phi=0$ and $R_m=1$.
The statements below hold in any model where CP is a good symmetry in the
relevant processes.

It is important to realize that the choice of $\phi=0$ is equivalent to
choosing $|D_1\rangle$ ($|D_2\rangle$) to be the CP-odd (CP-even) state,
$|D_-\rangle$ ($|D_+\rangle$). This can be seen from eq. \parlkpi.
It gives $\lambda_{K^+K^-}=-1$. We define the CP-odd state as the
mass eigenstate that does not decay into $K^+K^-$. Indeed, we now have
\eqn\defodd{
\vev{K^+K^-|{\cal H}|D_1}\ =\ pA_{K^+K^-}(1+\lambda_{K^+K^-})=0.}

In the CP limit, a non-zero value of $y_{\rm CP}$ (see eq. \ycpnum)
requires unambiguously that the width difference is large:
\eqn\ycpsm{y={\Gamma_+-\Gamma_-\over2\Gamma}=(3.42\pm1.39\pm0.74)\times10^{-2}.}
The fact that $y>0$ is preferred suggests that the CP-even state has a shorter lifetime,
that is $|D_{+,-}\rangle=|D_{S,L}\rangle$ where $S$ and $L$ stands for
`short' and `long' lifetimes, respectively. This important result holds
in the CP limit model independently.

\newsec{The Case of $|M_{12}/\Gamma_{12}|\ll1$}
It could be the case that $SU(3)$ breaking effects are stronger for the
absorptive part of the $\DDbar$ transition amplitude, $\Gamma_{12}$, than for the
dispersive part, $M_{12}$. In this section we investigate the implications of the
FOCUS and CLEO results in case that indeed
\eqn\smallmot{|M_{12}/\Gamma_{12}|\ll1.}
When we neglect small effects of ${\cal O}(|M_{12}/\Gamma_{12}|)$, several simplifications
occur. Define 
\eqn\defpot{\phi_{12}\equiv\arg(M_{12}/\Gamma_{12}).} 
Then, to leading order in $|M_{12}/\Gamma_{12}|$, we have:
\eqn\smallxy{\eqalign{
{x/ y}\ =&\ 2\left|{M_{12}/\Gamma_{12}}\right|\cos\phi_{12},\cr
A_m\ =&\ 4\left|{M_{12}/\Gamma_{12}}\right|\sin\phi_{12},\cr
\phi\ =&\ -2\left|{M_{12}/\Gamma_{12}}\right|^2\sin2\phi_{12}.\cr}}
We learn that in the limit \smallmot, $x$ can be neglected and all CP violating
effects can be neglected. This should be contrasted with the case of
$|\Gamma_{12}/M_{12}|\ll1$, which holds for the $B$ and $B_s$ mesons,
where the effects of $A_m$ can be neglected but those of $\phi$ are
not suppressed. There are two interesting consequnces of this difference. 
First, in the $B_s$ system, a lifetime difference between CP eigenstates 
and flavor specific final states (analoguous to $y_{\rm CP}$ of eq. \FkkGcpv)
measures $\Delta\Gamma(B_s)$ only if there is no new CP violation in the mixing
\ref\YBs{Y. Grossman, Phys. Lett. B380 (1996) 99, hep-ph/9603244.}.
In the $D$ system, if \smallmot\ holds, $y_{\rm CP}\approx y$
model independently. Second, {\it even in the case that new physics dominates 
$M_{12}(D)$, the sensitivity of any physical observable to it is suppressed 
by $|M_{12}/\Gamma_{12}|$.}

Neglecting $x$, $A_m$ and $\phi$, the FOCUS and CLEO results can be written as follows:
\eqn\smallmexp{\eqalign{
y\ =&\ (3.42\pm1.57)\times10^{-2},\cr
y\cos\delta\ =&\ (-2.5^{+1.4}_{-1.6})\times10^{-2},\cr
y\sin\delta\ =&\ (0.0\pm1.5)\times10^{-2}.\cr}}
The FOCUS measurement determines directly $y$. The first two equations give
\eqn\smallmdel{\cos\delta=-0.73^{+0.55}_{-0.27}.}
The third equation requires that $|\sin\delta|$ is not large and
consequently narrows the range for $\delta$ even further,
\eqn\smadel{\cos\delta\lsim-0.5.}
 
The conclusion of our discussion here is that if the $\DDbar$ system
provides a (unique!) example of $|M_{12}|\ll|\Gamma_{12}|$, then
the FOCUS and CLEO measurements determine $y$ to be at the few percent
level and the strong phase $\delta$ is well above $\pi/2$. 

\newsec{Implications for the Width Difference}
The value of the phase $\delta$ has important implications for another
aspect of our study, that is the width difference. The contributions of
the four charged two-body states,
\eqn\foumod{n_{2c}\ =\ K^+K^-,\ \pi^+\pi^-,\ K^+\pi^-,\ K^-\pi^+,}
to $\Gamma_{12}$, the absorptive part of the transition amplitude 
$\vev{D^0|{\cal H}|\barD}$, can be written as
\eqn\gamcha{
(\Gamma_{12})_{2c}\ =\ \sum_{n_{2c}} A_{n_{2c}}^*\bar A_{n_{2c}},}
which leads to the following contribution to $y$:
\eqn\ycharged{y_{2c}=\ \BR(D^0\rightarrow K^-K^+)
+\BR(D^0\rightarrow \pi^-\pi^+)-2\cos\delta\sqrt{R}\
\BR(D^0\rightarrow K^+\pi^-).}
There are two points that we would like to extract from eq. \ycharged.
First, in the $SU(3)$ limit, $\BR(D^0\rightarrow K^-K^+)=
\BR(D^0\rightarrow \pi^-\pi^+)=\sqrt{R}\ \BR(D^0\rightarrow K^+\pi^-)$. 
The phase $\delta$ defined in \parlkpi\ vanishes in the $SU(3)$ limit
which is consitent with the fact that $y_{2c} = 0$ in this limit.
Second, we can use the measured
branching ratios for the four decay modes and the value of the phase $\delta$
as fitted to the CLEO and FOCUS results to estimate $y_{2c}$. We use 
\ref\PDG{C. Caso {\it et al.}, The Eur. Phys. J. C3 (1998) 1, 
and 1999 off-year partial update for the 2000 edition available on 
the PDG WWW pages (URL: http://pdg.lbl.gov/).}\
\eqn\brbkpi{\eqalign{
\BR(D^0\rightarrow K^-\pi^+)\ =&\ (3.83\pm0.09)\times10^{-2},\cr
\BR(D^0\rightarrow \pi^-\pi^+)\ =&\ (1.52\pm0.09)\times10^{-3},\cr
\BR(D^0\rightarrow K^-K^+)\ =&\ (4.24\pm0.16)\times10^{-3},\cr}}
and \cleoexp\ (see eq. \cleoyx)
\eqn\sqrtr{\sqrt{R}=0.069\pm0.009.}
Using central values for the branching ratios, we get:
\eqn\ychnu{y_{2c}\sim(5.76-5.29\cos\delta)\times10^{-3}.}
Taking $-1\lsim\cos\delta\lsim0$ from \delxy, we find
\eqn\ychfin{0.6\times10^{-2}\lsim y_{2c}\lsim1.1\times10^{-2},}
to be compared with the range \ycpsm\ for $y$. Note that the sign
of this contribution is consistent with the overall sign of $y$ as
measured by FOCUS. There are of course other
intermediate states that contribute to $y$. Eq. \ychfin\ suggests that,
if the strong phases strongly violate $SU(3)$ as required for consistency of
the CLEO and FOCUS results, such contributions could easily be at the percent 
level as required by the same experiments.

\newsec{Conclusions}
The FOCUS and CLEO collaborations have provided new measurements of
the $\DDbar$ mixing parameters that are sensitive to effects of order
a few percent. FOCUS obtains a 2.2$\sigma$ signal and CLEO obtains a
1.8$\sigma$ signal of such effects. It could well be that these signals
are just statistical fluctuations and that the mixing parameters
are much smaller than the percent level. This is the theoretical wisdom,
based on the Standard Model and on approximate flavor $SU(3)$.
If, however, the central values of the two measurements are close
to the true values, then at least the assumption of approximate $SU(3)$
for the strong interactions has to be modified. In particular, there are two 
{\it independent} pieces of evidence that the strong phase in 
$D\rightarrow K^\pm\pi^\mp$ decays is very large, $\delta\gsim\pi/4$ and perhaps 
$\delta\sim3\pi/4$ (while $\delta=0$ in the $SU(3)$ limit):
\item{(i)} Either a negative sign for $\cos\delta$ or large $x$ and large 
$\sin\delta$ are necessary to make the signs of the mixing parameters measured 
by FOCUS and by CLEO consistent with each other.
\item{(ii)} $\cos\delta$ far from its $SU(3)$ limit value of one implies 
that some contributions to the width difference are at the percent level.

We also discussed the possibility that in the $\DDbar$ system $|M_{12}/\Gamma_{12}|\ll1$, 
in contrast to the neutral $B$ meson systems. In such a case, the $\DDbar$ system
is not sensitive to new physics, even if new physics dominates $M_{12}$.
In particular, CP is expected to be a good symmetry regardless of whether
there are large CP violating contributions to $M_{12}$. The above statements
about large $SU(3)$ breaking effects become even sharper in this case.

A much clearer picture would emerge if the accuracy of the
measurements improves and, in particular, if the mixing parameters are
measured separately in the $D^0$ and $\barD$  decays. For example, 
the FOCUS collaboration has summed over the $D^0\rightarrow K^+K^-$ and 
$\barD\rightarrow K^+K^-$ modes, but there is much to learn from comparing 
them to each other. Explicitly, we obtain from eq. \fitexp:
\eqn\compexp{
{\hat\Gamma(D^0\rightarrow K^+K^-)-\hat\Gamma(\barD\rightarrow K^+K^-)\over
\hat\Gamma(D^0\rightarrow K^+K^-)+\hat\Gamma(\barD\rightarrow K^+K^-)}=
{A_m\over2}y\cos\phi-x\sin\phi.}
A difference between the fitted decay width of the two CP conjugate modes
will provide important information on the CP violating parameters.
  
\vskip 3cm

\centerline{\bf Acknowledgments}
\noindent We thank David Asner, Adam Falk, Harry Nelson and Jim Wiss
for useful discussions.
S.B. thanks the School of Natural Sciences at the Institute for 
Advanced Study for the hospitality.
Y.G. is supported by the U.S. Department of Energy under
contract DE-AC03-76SF00515.
Fermilab is operated by Universities Research
Association, Inc., under DOE contract DE-AC02-76CH03000.
Y.N. is supported by the Department of Energy under contract
No.~DE--FG02--90ER40542, by the Ambrose Monell Foundation, 
by AMIAS (Association of Members of the Institute for Advanced Study),
by the Israel Science Foundation founded by the Israel Academy of Sciences
and Humanities, and by the Minerva Foundation (Munich).  

\listrefs
\end